\documentclass{article}

\usepackage{arxiv}

\usepackage[utf8]{inputenc} % allow utf-8 input
\usepackage[T1]{fontenc}    % use 8-bit T1 fonts
\usepackage{hyperref}       % hyperlinks
\usepackage{url}            % simple URL typesetting
\usepackage{booktabs}       % professional-quality tables
\usepackage{amsfonts}       % blackboard math symbols
\usepackage{nicefrac}       % compact symbols for 1/2, etc.
\usepackage{microtype}      % microtypography
\usepackage{lipsum}
\usepackage{graphicx}
\graphicspath{ {./images/} }
\usepackage{amsmath}
\usepackage{bm}
\usepackage{amssymb,amsthm,amsmath}
\usepackage{xcolor,paralist,hyperref,titlesec,fancyhdr,etoolbox}
\usepackage{algorithm}
\usepackage{algpseudocode}
\usepackage{makecell}

\title{Koopman Spectral Reduced-Order Modeling of Spherical Diffusion in Lithium-Ion Batteries}

\author{
 Jihoon Moon \\
  Department of Mechanical Engineering\\
  The Pennsylvania State University\\
  University Park, PA 16802 \\
  \texttt{jihoonmoon@psu.edu} \\
}

\begin{document}
\maketitle
\begin{abstract}
Physics-based battery models provide internal electrochemical states for estimation and control, but solving the partial differential equations governing solid diffusion can be computationally expensive. This paper therefore develops an analytical Koopman spectral reduced-order model for the single particle model. The Koopman eigenfunctionals and eigenvalues are derived directly from the eigenfunctions of the self-adjoint zero-flux diffusion operator. The zero mode represents the volume average concentration, the nonzero modes describe decaying radial gradients, and projection of the current dependent surface flux yields a linear state-space model that reconstructs the average, surface, and full radial concentrations. Unlike data-driven Koopman models, the proposed formulation requires neither training data nor empirical lifting functions. Compared with a 400 control volume finite volume model (FVM) during a constant 1C discharge, the 80 Koopman mode model achieves negative and positive surface concentration RMSE values of $1.26\times10^{-5}\,\mathrm{mol/m^{3}}$ and $4.10\times10^{-6}\,\mathrm{mol/m^{3}}$, a terminal voltage RMSE of $8.42\times10^{-4}\,\mathrm{V}$. The models with 5 and 80 Koopman modes were approximately 225 and 31 times faster than the FVM model, respectively. These results demonstrate a physically interpretable and computationally efficient representation of battery diffusion dynamics.
\end{abstract}

% keywords can be removed

\section{Introduction}

Lithium-ion batteries are widely used in electrified transportation, grid energy storage, and portable systems because of their high energy density and efficiency. Their safe and efficient operation relies on battery management systems (BMSs) that require models capable of predicting measurable outputs and internal electrochemical states. Equivalent circuit models (ECMs) are attractive for real-time applications because of their low computational cost and have been widely used for state estimation, health monitoring, and fault diagnosis \cite{ahuja2026lithium,moon2024short,moon2026detecting,moon2025safe,bhaskar2024post,moon2024state}. However, their parameters are largely empirical and provide limited information about internal concentration dynamics. Physics-based electrochemical models instead describe lithium transport and reaction mechanisms through conservation laws and constitutive relations, but their governing PDEs and differential-algebraic equations can require substantial computational effort. This tradeoff has motivated the development of reduced-order electrochemical models for real-time battery applications \cite{Tanim2014ACC,Tanim2015JDSMC,Moon2022IFAC}.

The single particle model (SPM) is a widely used physics-based simplification that represents each porous electrode by a representative spherical particle and describes solid-phase lithium transport using spherical diffusion. The SPM retains physically meaningful quantities, including volume average and surface concentrations, while substantially reducing the complexity of higher-order electrochemical models. Nevertheless, direct spatial discretization of the diffusion PDE introduces multiple states per particle, and higher spatial resolution is required to accurately capture surface concentration dynamics. Several reduced-order approaches have therefore been proposed. Tanim et al. developed low-order electrolyte enhanced SPM formulations in explicit transfer function form \cite{Tanim2014ACC,Tanim2015JDSMC}, while Docimo et al. derived a second-order physics-based battery model with an equivalent circuit interpretation \cite{Docimo2014DSCC}. Pade approximations have also been widely applied to spherical diffusion transfer functions to obtain low-order state-space models \cite{Forman2011JES}, and optimization-based reduced-order formulations have been developed and compared with Pade-based models, including extensions incorporating degradation effects \cite{Moon2022IFAC}. These approaches demonstrate that substantial model order reduction is possible while retaining electrochemical structure. However, most existing SPM reductions rely on rational approximation, spatial discretization, moment approximation, or parameter fitting rather than directly exploiting the analytical spectral structure of the spherical diffusion operator.

Koopman operator theory provides an alternative spectral framework in which dynamical systems are represented through the evolution of observable functions. Koopman-based reduced representations and control formulations have been developed for PDEs \cite{Kutz2018KoopmanPDE,Peitz2019PDE,Korda2018MPC}. In particular, Nakao and Mezi'{c} showed that, for linear diffusion PDEs, Koopman eigenfunctionals can be constructed as linear functionals of the field variable associated with eigenfunctions of the spatial operator \cite{Nakao2020KoopmanPDE}. This establishes a direct connection between classical diffusion eigenfunction expansions and Koopman spectral coordinates.

Koopman methods have recently been explored for lithium-ion battery modeling and estimation. Choi et al. developed a data-driven Koopman surrogate of an SPM using learned encoder and decoder mappings \cite{Choi2023KoopmanBattery}, while Gupta et al. combined physics-based features with streamed battery data in a physics-informed Koopman framework for state-of-charge estimation \cite{Gupta2025DPIK}. These approaches primarily identify the lifted coordinates and operators from data. In contrast, the spherical solid diffusion PDE of the SPM admits an analytical construction of Koopman eigenfunctionals directly from the governing diffusion operator and boundary conditions. This work develops a physics-based Koopman spectral reduced-order model in which the reduced coordinates, eigenvalues, and spatial modes are obtained analytically. The resulting representation is compact, physically interpretable, and suitable for real-time battery estimation and control.

This work develops a Koopman spectral reduced-order model of SPM with four principal contributions. First, the Koopman eigenfunctionals are derived analytically from the eigenfunctions of the self-adjoint diffusion operator. Second, the resulting coordinates are physically interpretable: the zero eigenfunctional recovers the volume average concentration, while the nonzero eigenfunctionals represent radial diffusion modes with analytical decay rates. Third, the autonomous and controlled dynamics are treated separately. The eigenfunctionals and eigenvalues are obtained from the zero-flux diffusion generator, and the current-dependent boundary flux is subsequently projected into these coordinates to form the controlled state-space model. Finally, the proposed model is compared with a high-order finite volume reference to assess concentration and voltage accuracy, and computational efficiency. This framework provides an analytical and interpretable alternative to learned Koopman embeddings and conventional rational approximations for control-oriented battery diffusion modeling.

\section{Methodology}\label{sec:methodology}

This section derives the Koopman spectral reduced-order model for solid-phase diffusion in the SPM. First, the spherical diffusion equations, boundary conditions, and modeling assumptions are presented. The eigenfunctions of the self-adjoint zero-flux diffusion operator are then used to define the Koopman eigenfunctionals and modal coordinates \cite{Nakao2020KoopmanPDE}. Next, the surface-flux input is projected onto these coordinates to obtain the boundary-controlled state-space model, followed by its exact discrete-time formulation. Finally, the numerical procedure used to compare the Koopman model with the finite-volume reference is described.

\subsection{Single-Particle Solid Diffusion Model}\label{subsec:spm}

The SPM represents each porous electrode by a single spherical active-material particle. The particles within an electrode are assumed to have identical radii and to experience a spatially uniform reaction rate. Electrolyte concentration gradients are neglected in the present solid-diffusion reduction. For electrode $i\in\{neg,pos\}$, where $neg$ and $pos$ denote the negative and positive electrodes, respectively, conservation of lithium in the solid particle is governed by Fick's law of diffusion,
\begin{equation}
\frac{\partial c_{i}(r,t)}{\partial t}
=
\frac{D_{i}}{r^2}
\frac{\partial}{\partial r}
\left(
 r^2\frac{\partial c_{i}(r,t)}{\partial r}
\right),
\qquad 0 \leq r \leq R_{i},
\label{eq:spherical_diffusion}
\end{equation}
where $c_{i}$ is the solid-phase lithium concentration, $D_{i}$ is the solid-phase diffusion coefficient, and $R_{i}$ is the particle radius. Spherical symmetry gives the center boundary condition and the electrochemical reaction at the particle surface as 
\begin{equation}
\left.\frac{\partial c_{i}}{\partial r}\right|_{r=0}=0, \qquad -D_{i}\left.\frac{\partial c_{i}}{\partial r}\right|_{r=R_{i}}
=J_i(t)
\label{eq:bc}
\end{equation}
where $J_i$ is the molar flux at the particle surface. Positive cell current is defined as discharge in this work. The electrode fluxes are written as
\begin{equation}
J_{neg}(t)=\frac{I(t)}{F A_{neg}a_{neg}L_{neg}},
J_{pos}(t)=-\frac{I(t)}{F A_{pos}a_{pos}L_{pos}},
\label{eq:current_flux}
\end{equation}
where $F$ is Faraday's constant, $A_{i}$ is the electrode area, $L_i$ is the electrode thickness, $a_{i}=\frac{3\varepsilon_{i}}{R_{i}}$ is the specific interfacial area, and $\varepsilon_{i}$ is the active-material volume fraction. The sign difference in \eqref{eq:current_flux} reflects lithium leaving the negative particle and entering the positive particle during discharge.

Two quantities of particular interest are the volume-average concentration and the surface concentration,
\begin{equation}
\bar c_{i}(t)
=
\frac{3}{R_{i}^3}
\int_0^{R_{i}}c_{i}(r,t)r^2\,dr,
\label{eq:average_concentration}
\end{equation}
\begin{equation}
c_{s,i}(t)=c_{i}(R_{i},t).
\label{eq:surface_concentration}
\end{equation}
The average concentration is directly related to the lithium inventory and electrode state of charge, and the surface concentration enters the overpotential and open circuit potential of the SPM. The overpotential and terminal voltage are presented as 
\begin{equation}
  \begin{aligned}
    &\eta_i = \frac{2R_uT}{F}\sinh^{-1}\left(\frac{J_i(t)}{2a_ik_i\left(c_e c_{s,i}\left(c_{max,i}-c_{s,i}\right)^{0.5}\right)}\right), \\
    &V(t) = U_{ref,pos}(c_{s,pos}) - U_{ref,neg}(c_{s,neg}) + \eta_{pos} - \eta_{neg} \\ &- \left( \left(\frac{R_{film,pos}}{A_{pos}L_{pos}a_{pos}}\right) + \left(\frac{R_{film,neg}}{A_{neg}L_{neg}a_{neg}}\right) + R_c \right)I(t)
  \end{aligned}\label{outputV}
\end{equation}
See \cite{Moon2022IFAC} for values of model parameters and open circuit potential equations.

\subsection{Koopman Formulation of the Spherical Diffusion Dynamics}\label{subsec:koopman_formulation}

The Koopman spectral representation developed in this work is constructed from the spatial eigenfunctions of the spherical diffusion operator. To determine these eigenfunctions, the autonomous diffusion dynamics are considered first by imposing the homogeneous surface condition $J_i(t)=0$. Let $S_i^t$ denote the evolution operator generated by \eqref{eq:spherical_diffusion}--\eqref{eq:bc} with a homogeneous surface boundary condition. For an observable functional $g[c]$, the Koopman operator is defined by
\begin{equation}
(\mathcal{K}_i^t g)[c_0]
=
g[S_i^t c_0],
\label{eq:koopman_definition}
\end{equation}
where $c_0(r)$ is an initial concentration profile. The associated infinitesimal generator is
\begin{equation}
\mathcal{L}_i g
=
\lim_{t\rightarrow 0}
\frac{\mathcal{K}_i^t g-g}{t}.
\label{eq:koopman_generator}
\end{equation}
A Koopman eigenfunctional $\psi_{i,n}$ satisfies
\begin{equation}
\mathcal{K}_i^t\psi_{i,n}
=
e^{\lambda_{i,n}t}\psi_{i,n},
\label{eq:koopman_eigenfunctional}
\end{equation}
or equivalently $\mathcal{L}_i\psi_{i,n}=\lambda_{i,n}\psi_{i,n}$. Thus, if the lifted coordinate is defined as $z_{i,n}(t)=\psi_{i,n}[c_{s,i}(\cdot,t)]$, its zero-input dynamics are
\begin{equation}
\dot z_{i,n}(t)=\lambda_{i,n}z_{i,n}(t).
\label{eq:zero_input_modal_dynamics}
\end{equation}

\subsubsection{Spherical Diffusion Eigenfunctions}
\label{subsubsec:spherical_eigenfunctions}

The spatial eigenfunctions are obtained from the autonomous diffusion dynamics by setting the surface flux to zero, $J_i(t)=0$. Assuming a separated solution,
\begin{equation}
c_i(r,t)=\phi_i(r)T_i(t),
\label{eq:separated_solution}
\end{equation}
and substituting \eqref{eq:separated_solution} into \eqref{eq:spherical_diffusion} gives
\begin{equation}
\frac{1}{D_iT_i}\frac{dT_i}{dt}
=
\frac{1}{\phi_i r^2}
\frac{d}{dr}
\left(
r^2\frac{d\phi_i}{dr}
\right).
\label{eq:separation}
\end{equation}
Since the left- and right-hand sides depend only on $t$ and $r$, respectively, both are set equal to the separation constant $-\mu^2$. The spatial eigenvalue problem is therefore
\begin{equation}
\frac{1}{r^2}
\frac{d}{dr}
\left(
r^2\frac{d\phi}{dr}
\right)
=
-\mu^2\phi,
\label{eq:spatial_eigenproblem}
\end{equation}
subject to
\begin{equation}
\phi'(0)=0,
\qquad
\phi'(R_i)=0.
\label{eq:eigen_bc}
\end{equation}
The corresponding temporal component satisfies
$T_i(t)=T_i(0)e^{-D_i\mu^2t}$, showing that the nonuniform diffusion modes decay exponentially.

Expanding \eqref{eq:spatial_eigenproblem} gives
\begin{equation}
\phi''+\frac{2}{r}\phi'+\mu^2\phi=0.
\label{eq:spherical_ode}
\end{equation}
Using the transformation $y=r\phi$ reduces this equation to
$y''+\mu^2y=0$, with solution
$y=A\sin(\mu r)+B\cos(\mu r)$. Requiring the concentration to remain finite at the particle center gives $B=0$. Choosing $A=1/\mu$ normalizes the eigenfunction such that $\phi(0)=1$, yielding
\begin{equation}
\phi_{i,n}(r)
=
\frac{\sin(\mu_{i,n}r)}
{\mu_{i,n}r},
\qquad n\geq1.
\label{eq:spherical_eigenfunction}
\end{equation}

Applying the surface condition $\phi_{i,n}'(R_i)=0$ gives
\begin{equation}
\mu_{i,n}R_i\cos(\mu_{i,n}R_i)
-
\sin(\mu_{i,n}R_i)
=
0.
\label{eq:root_condition_dimensional}
\end{equation}
Defining $\zeta_n=\mu_{i,n}R_i$ gives $\tan\zeta_n=\zeta_n$.
The same dimensionless roots $\zeta_n$ therefore apply to both electrodes. For $\mu=0$, the eigenvalue problem admits the constant mode $\phi_{i,0}(r)=1$.
The functions $\{\phi_{i,n}\}_{n=0}^{\infty}$ define the intrinsic spatial modes of the autonomous spherical diffusion dynamics.

\subsubsection{Koopman Eigenfunctionals and Modal Coordinates}
\label{subsubsec:koopman_modal_coordinates}

The connection between the spatial diffusion modes and the Koopman eigenfunctionals follows from the spectral properties of the spherical diffusion operator. Define
\begin{equation}
\mathcal{D}_i f
=
\frac{1}{r^2}
\frac{d}{dr}
\left(
r^2\frac{df}{dr}
\right),
\label{eq:spatial_operator}
\end{equation}
with the spherical weighted inner product
\begin{equation}
\langle f,g\rangle_i
=
\int_0^{R_i}
f(r)g(r)r^2\,dr.
\label{eq:weighted_inner_product}
\end{equation}
The factor $r^2$ originates from the spherical volume element $dV=4\pi r^2dr$.

An operator is self-adjoint when
$\langle f,\mathcal{D}_i g\rangle_i
=
\langle\mathcal{D}_i f,g\rangle_i$.
For functions satisfying the homogeneous Neumann boundary conditions, integration by parts gives
\begin{align}
\langle f,\mathcal{D}_i g\rangle_i =
\left[fr^2g'\right]_0^{R_i}
-
\int_0^{R_i}r^2f'g'\,dr \nonumber =
-\int_0^{R_i}r^2f'g'\,dr
=
\langle\mathcal{D}_i f,g\rangle_i.
\label{eq:self_adjoint}
\end{align}
The boundary term vanishes because $g'(R_i)=0$ at the particle surface and $r^2=0$ at the center. Hence, $\mathcal{D}_i$ is self-adjoint under \eqref{eq:weighted_inner_product}.

Self-adjointness is important for the Koopman construction in two ways. First, eigenfunctions corresponding to distinct eigenvalues are orthogonal. Using $\mathcal{D}_i\phi_{i,n}=-\mu_{i,n}^2\phi_{i,n}$, together with self-adjointness gives
\begin{equation}
\left(
\mu_{i,m}^2-\mu_{i,n}^2
\right)
\langle\phi_{i,m},\phi_{i,n}\rangle_i
=
0.
\end{equation}
Therefore, for $m\neq n$, $\langle\phi_{i,m},\phi_{i,n}\rangle_i=0$.
Together with the completeness of this eigenfunction set, the concentration field can be expanded as
\begin{equation}
c_i(r,t)
=
\sum_{n=0}^{\infty}
z_{i,n}(t)\phi_{i,n}(r).
\label{eq:modal_expansion}
\end{equation}

The modal coefficient is obtained by projecting \eqref{eq:modal_expansion} onto $\phi_{i,n}$. Defining
\begin{equation}
N_{i,n}
=
\langle\phi_{i,n},\phi_{i,n}\rangle_i
=
\int_0^{R_i}
\phi_{i,n}^2(r)r^2\,dr,
\label{eq:normalization_definition}
\end{equation}
orthogonality gives
\begin{equation}
z_{i,n}(t)
=
\frac{1}{N_{i,n}}
\int_0^{R_i}
c_i(r,t)\phi_{i,n}(r)r^2\,dr.
\label{eq:modal_coordinate}
\end{equation}
For the zero mode,
\begin{equation}
N_{i,0}=\frac{R_i^3}{3},
\qquad
z_{i,0}(t)=\bar c_i(t),
\label{eq:zero_mode_normalization}
\end{equation}
so the zero modal coordinate is exactly the volume-average concentration. For $n\geq1$, substitution of \eqref{eq:spherical_eigenfunction} into \eqref{eq:normalization_definition}, together with $\tan\zeta_n=\zeta_n$, gives
\begin{equation}
N_{i,n}
=
\frac{R_i^3}{2(1+\zeta_n^2)}.
\label{eq:normalization_closed}
\end{equation}

The second role of self-adjointness is to establish the Koopman eigenfunctionals. For a general linear PDE, Koopman eigenfunctionals are constructed from eigenfunctions of the adjoint spatial operator \cite{Nakao2020KoopmanPDE}. Since $\mathcal{D}_i$ is self-adjoint, its adjoint eigenfunctions coincide with $\phi_{i,n}$. Accordingly, define
\begin{equation}
\psi_{i,n}[c]
=
\frac{1}{N_{i,n}}
\langle c,\phi_{i,n}\rangle_i.
\label{eq:koopman_observable}
\end{equation}
Along the autonomous zero-flux diffusion dynamics,
\begin{align}
\frac{d}{dt}\psi_{i,n}[c_i]
&=
\frac{D_i}{N_{i,n}}
\left\langle
\mathcal{D}_i c_i,\phi_{i,n}
\right\rangle_i =
\frac{D_i}{N_{i,n}}
\left\langle
c_i,\mathcal{D}_i\phi_{i,n}
\right\rangle_i \nonumber =
-D_i\mu_{i,n}^2\psi_{i,n}[c_i].
\label{eq:eigenfunctional_proof}
\end{align}
Thus, the Koopman eigenvalues are
\begin{equation}
\lambda_{i,0}=0,
\qquad
\lambda_{i,n}
=
-D_i
\left(
\frac{\zeta_n}{R_i}
\right)^2,
\quad n\geq1,
\label{eq:koopman_eigenvalues}
\end{equation}
and
\begin{equation}
\mathcal{K}_i^t\psi_{i,n}
=
e^{\lambda_{i,n}t}\psi_{i,n}.
\end{equation}
Hence, $\psi_{i,n}$ is a Koopman eigenfunctional and $z_{i,n}(t)=\psi_{i,n}[c_i(\cdot,t)]$ is its corresponding Koopman coordinate. Here, $\phi_{i,n}(r)$ specifies the spatial shape of the $n$th diffusion mode, while $z_{i,n}(t)$ specifies its time-dependent amplitude.

The eigenfunctionals and eigenvalues above characterize the autonomous zero-flux diffusion generator. The current-dependent nonhomogeneous surface flux is subsequently projected onto these Koopman coordinates to obtain the boundary-controlled reduced-order model.

\subsection{Boundary-Controlled Koopman Reduced-Order Model}\label{subsec:controlled_koopman}

The eigenfunctions above are obtained from the homogeneous diffusion operator, while battery current enters through the nonhomogeneous surface-flux condition \eqref{eq:bc}. The effect of the boundary input on each Koopman coordinate is obtained by projecting the full PDE onto $\phi_{i,n}$. Multiplying \eqref{eq:spherical_diffusion} by $\phi_{i,n}r^2$ and integrating gives
\begin{equation}
N_{i,n}\dot z_{i,n}
=
D_{s,i}
\int_0^{R_{i}}
\phi_{i,n}
\frac{d}{dr}\left(r^2c_{i,r}\right)dr,
\label{eq:input_projection_start}
\end{equation}
where $c_{i,r}=\frac{\partial c_{s,i}}{\partial r}$. Integration by parts gives
\begin{align}
N_{i,n}\dot z_{i,n}
=&
D_{i}\left[\phi_{i,n}r^2c_{i,r}\right]_0^{R_{i}}
-
D_{i}\int_0^{R_{i}}\phi_{i,n}'r^2c_{i,r}\,dr  =
D_{i}R_{i}^2\phi_{i,n}(R_{i})c_{i,r}(R_{i},t) + D_{i}\int_0^{R_{i}}
c_{i}\frac{d}{dr}\left(r^2\phi_{i,n}'\right)dr.
\label{eq:input_projection_ibp}
\end{align}
The boundary term generated by the second integration by parts vanishes because $r^2=0$ at the center and $\phi_{i,n}'(R_{i})=0$ at the surface. Using the eigenvalue equation and boundary condition at the particle surface in \eqref{eq:bc}, \eqref{eq:input_projection_ibp} becomes
\begin{equation}
N_{i,n}\dot z_{i,n}
=
-D_{i}\mu_{i,n}^2N_{i,n}z_{i,n}
-
R_{i}^2\phi_{i,n}(R_{i})J_i(t).
\label{eq:controlled_modal_before_divide}
\end{equation}
Therefore,
\begin{equation}
\dot z_{i,n}=\lambda_{i,n}z_{i,n}+b_{i,n}J_i(t),
\label{eq:controlled_modal}
\end{equation}
where
\begin{equation}
b_{i,n}=-\frac{R_{i}^2\phi_{i,n}(R_{i})}{N_{i,n}}.
\label{eq:b_general}
\end{equation}
For the zero mode,
\begin{equation}
b_{i,0}=-\frac{3}{R_{i}},
\qquad
\dot{\bar c}_{i}=-\frac{3}{R_{i}}J_i,
\label{eq:zero_mode_controlled}
\end{equation}
For $n\geq1$, the root condition gives $\phi_{i,n}(R_{i})=\cos\zeta_n$, and \eqref{eq:b_general} can be written as
\begin{equation}
b_{i,n}
=
-\frac{2(1+\zeta_n^2)\cos\zeta_n}{R_{i}}.
\label{eq:b_closed}
\end{equation}

Retaining $N_K$ nonzero modes for electrode $i$ gives the finite-dimensional state
\begin{equation}
\bm{z}_i
=
\begin{bmatrix}
z_{i,0} & z_{i,1} & \cdots & z_{i,N_K}
\end{bmatrix}^{\!T},
\label{eq:electrode_state}
\end{equation}
and the continuous-time reduced model
\begin{equation}
\dot{\bm{z}}_i
=
\bm{A}_i\bm{z}_i+\bm{B}_{J,i}J_i,
\label{eq:electrode_state_space}
\end{equation}
with
\begin{equation}
\begin{aligned}
  \bm{A}_i =
\operatorname{diag}
\left(\lambda_{i,0},\lambda_{i,1},\ldots,\lambda_{i,N_K}\right), \\
\bm{B}_{J,i}
=
\begin{bmatrix}
b_{i,0} & b_{i,1} & \cdots & b_{i,N_K}
\end{bmatrix}^{\!T}.
\end{aligned}
\label{eq:electrode_AB}
\end{equation}
The full radial concentration profile is reconstructed from
\begin{equation}
\begin{aligned}
  \hat c_{i}(r,t)&=\bm{\Phi}_i(r)\bm{z}_i(t), \\
\bm{\Phi}_i(r)&=
\begin{bmatrix}
\phi_{i,0}(r) & \phi_{i,1}(r) & \cdots & \phi_{i,N_K}(r)
\end{bmatrix}.
\end{aligned}
\label{eq:profile_reconstruction}
\end{equation}
The average and surface concentrations are linear outputs,
\begin{equation}
\bar c_{i}=\bm{C}_{\mathrm{avg},i}\bm{z}_i,
\qquad
c_{s,i}=\bm{C}_{s,i}\bm{z}_i,
\label{eq:modal_outputs}
\end{equation}
where
\begin{equation}
\begin{aligned}
\bm{C}_{\mathrm{avg},i}
&=
\begin{bmatrix}1&0&\cdots&0\end{bmatrix},
\\
\bm{C}_{s,i}
&=
\begin{bmatrix}
1 & \cos\zeta_1 & \cdots & \cos\zeta_{N_K}
\end{bmatrix}.
\end{aligned}
\label{eq:output_matrices}
\end{equation}

For the complete SPM diffusion dynamics, define
\begin{equation}
\bm{z}
=
\begin{bmatrix}
\bm{z}_{neg}^T & \bm{z}_{pos}^T
\end{bmatrix}^{T}.
\label{eq:combined_state}
\end{equation}
Using \eqref{eq:current_flux}, the current-driven model is
\begin{equation}
\dot{\bm{z}}
=
\bm{A}\bm{z}+\bm{B}_I I,
\label{eq:combined_continuous}
\end{equation}
where
\begin{equation}
\bm{A}=\operatorname{blkdiag}(\bm{A}_{neg},\bm{A}_{pos}),
\quad
\bm{B}_I
=
\begin{bmatrix}
\gamma_{neg}\bm{B}_{J,neg}\\
\gamma_{pos}\bm{B}_{J,pos}
\end{bmatrix},
\label{eq:combined_AB}
\end{equation}
with
\begin{equation}
\gamma_{neg}=\frac{1}{F A_{neg}a_{neg}L_{neg}},
\gamma_{pos}=-\frac{1}{F A_{pos}a_{pos}L_{pos}}.
\label{eq:gamma_definition}
\end{equation}

\section{Results and Discussion}
\subsection{Numerical Comparison Procedure}\label{subsec:fvm}

A finite volume method (FVM) with $N_{\mathrm{FVM}}=400$ control volumes is used as the high-order numerical reference. The FVM and Koopman models solve the same spherical diffusion problem with identical parameters, initial conditions, and surface-flux inputs; details of the FVM discretization are provided in \cite{xu2024finite}.

The number of nonzero Koopman modes is varied as $N_K\in\{5,10,20,40,80\}$ for both electrodes. The surface concentrations obtained from \eqref{eq:modal_outputs} are used to calculate the terminal voltage in \eqref{outputV}. Model performance is evaluated over a constant 1C discharge from $100\%$ to $0\%$ state of charge (SOC).

Accuracy is evaluated for the surface concentration and terminal voltage time histories and for the radial concentration profiles. For a sampled output $y(t)$, the time-domain root mean square error (RMSE) is
\begin{equation}
\mathrm{RMSE}_{t}(y)
=
\sqrt{\frac{1}{N_t}\sum_{k=1}^{N_t}
\left[y^{\mathrm{K}}(t_k)-y^{\mathrm{FVM}}(t_k)\right]^2},
\label{eq:rmse_time}
\end{equation}
where $N_t=3600$ is the number of temporal samples. At a specified time $t$, the radial profile error for electrode $i$ is
\begin{equation}
\mathrm{RMSE}_{r,i}(t)
=
\sqrt{\frac{1}{N_{\mathrm{FVM}}}\sum_{j=1}^{N_{\mathrm{FVM}}}
\left[\hat c_i^{\mathrm{K}}(r_j,t)-c_i^{\mathrm{FVM}}(r_j,t)\right]^2},
\label{eq:rmse_radial}
\end{equation}
where $r_j$ denotes the center of the $j$th FVM control volume. 
Simulation runtime is also recorded to quantify computational cost. Together, these metrics assess temporal accuracy, spatial reconstruction, and the tradeoff between modal order and computational efficiency.

\begin{figure}
    \centering
    \includegraphics[width=100mm]{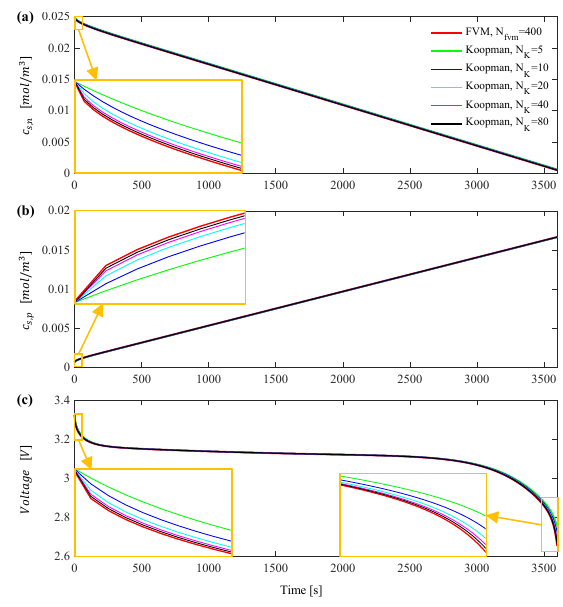}
    \caption{Surface concentration and terminal voltage responses predicted by the FVM and Koopman reduced-order models during a constant 1C discharge: (a) negative electrode surface concentration, (b) positive electrode surface concentration, and (c) terminal voltage.}
    \label{figure_2}
\end{figure}

\subsection{Comparison Results}
Figure \ref{figure_2} compares the negative and positive electrode surface concentrations and terminal voltage predicted by the FVM and Koopman models during the 1C discharge from $100\%$ to $0\%$ SoC. Every modal order captures the overall concentration and voltage trends. The enlarged plots make the remaining differences easier to observe. The low-order models show their largest surface concentration errors immediately after the discharge begins. A visible voltage error also develops near the end of discharge. These differences decrease as $N_K$ increases because the additional modes capture the faster diffusion dynamics. For $N_K=80$, the Koopman and FVM curves are nearly indistinguishable. The corresponding time-domain RMSE values ($\mathrm{RMSE_t}$) for the negative electrode surface concentration, positive electrode surface concentration, and terminal voltage are $1.26\times10^{-5}\,\mathrm{mol\,m^{-3}}$, $4.10\times10^{-6}\,\mathrm{mol\,m^{-3}}$, and $8.42\times10^{-4}\,\mathrm{V}$, respectively. Thus, the analytically derived Koopman coordinates accurately predict the surface concentration and the terminal voltage.

\begin{figure*}
    \centering
    \includegraphics[width=\linewidth]{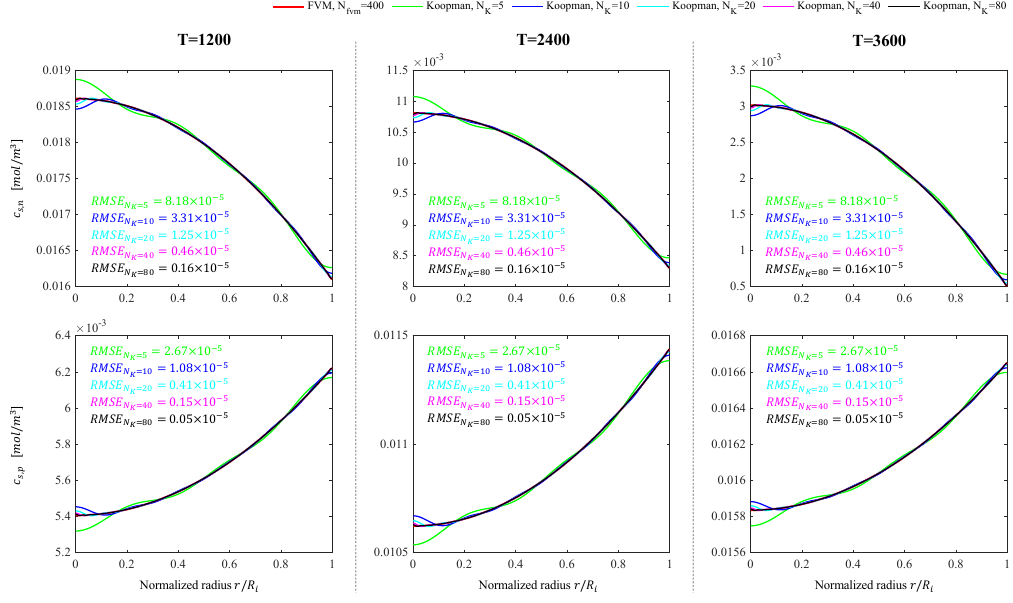}
    \caption{Radial concentration profiles predicted by the FVM and Koopman reduced-order models during a constant 1C discharge at $t=1200$, $2400$, and $3600\,\mathrm{s}$: negative electrode (top row) and positive electrode (bottom row).}
    \label{figure_3}
\end{figure*}

Figure \ref{figure_3} compares the radial concentration profiles at $t=1200$, $2400$, and $3600\,\mathrm{s}$. The top and bottom rows show the negative and positive electrodes, respectively, and the profiles are obtained directly from the Koopman coordinates using \eqref{eq:profile_reconstruction}. At all three snapshots, the largest deviations for $N_K=5$ and $N_K=10$ occur near the particle centers; increasing the modal order progressively removes these localized errors. The annotated negative-electrode radial RMSE ($\mathrm{RMSE_r}$) decreases from $8.18\times10^{-5}\,\mathrm{mol/m^{3}}$ for $N_K=5$ and $3.31\times10^{-5}\,\mathrm{mol/m^{3}}$ for $N_K=10$ to $0.16\times10^{-5}\,\mathrm{mol/m^{3}}$ for $N_K=80$. For the positive electrode, the corresponding values decrease from $2.67\times10^{-5}$ and $1.08\times10^{-5}\,\mathrm{mol/m^{3}}$ to $0.05\times10^{-5}\,\mathrm{mol/m^{3}}$. The RMSE decreases consistently as $N_K$ increases. This trend shows that model accuracy can be improved systematically by increasing the modal order. Moreover, unlike Pade approximation approach, the proposed spectral representation reconstructs the complete internal concentration field directly from the same physically interpretable coordinates.

The FVM requires $0.27\,\mathrm{s}$ to simulate the full discharge, whereas the Koopman models require $0.0088\,\mathrm{s}$ for $N_K=80$ and $0.0012\,\mathrm{s}$ for $N_K=5$. These values correspond to speedups of approximately $31$ and $225$, respectively. Because the high-order Koopman model is nearly indistinguishable from the FVM and the low-order models retain small concentration errors, the modal order can be selected according to the required balance between accuracy and computational cost. The combination of analytical construction, full field reconstruction, systematic modal convergence, and reduced runtime demonstrates the principal advantage of the proposed model for real-time battery estimation and control.

\section{Conclusion}

This work developed an analytical Koopman spectral reduced-order model for spherical solid-phase diffusion in the lithium-ion battery SPM. Eigenfunctionals, eigenvalues, and spatial modes were derived directly from the self-adjoint diffusion operator, with the zero coordinate representing the volume-average concentration and the nonzero coordinates describing radial gradients. Projecting the current dependent surface flux onto these coordinates produced a linear state-space model that reconstructs the average, surface, and full radial concentrations. Comparison with a 400 control volume FVM showed systematic convergence with increasing modal order, and the 80 Koopman mode model achieved surface concentration RMSE values of $1.26\times10^{-5}\,\mathrm{mol\,m^{-3}}$ and $4.10\times10^{-6}\,\mathrm{mol\,m^{-3}}$ for the negative and positive electrodes, respectively, with a terminal-voltage RMSE of $8.42\times10^{-4}\,\mathrm{V}$. The 80 and 5 Koopman mode models were approximately 31 and 225 times faster than the FVM, demonstrating an explicit accuracy-complexity tradeoff. These results establish the proposed model as a physically interpretable and computationally efficient framework for real-time battery estimation and control. Future work will address nonlinear transport properties, experimental validation under dynamic current profiles, and integration with estimator and controller frameworks.

\nocite{*}
\bibliographystyle{IEEEtran}
\bibliography{IEEEfull}

\end{document}